\documentclass[5p,twocolumn]{elsarticle}
\usepackage{lineno,hyperref}
\modulolinenumbers[5]
\usepackage{graphicx}
\usepackage{dcolumn}
\usepackage{bm}
\usepackage{amsmath}

\begin{document}
\begin{frontmatter}
  
  \title{Self-consistent calculations of electron-capture decays in Z=118,
    119, and 120 superheavy isotopes}

\author{P.~Sarriguren}
\address{Instituto de Estructura de la Materia, IEM-CSIC, Serrano 123,
  E-28006 Madrid, Spain}
\ead{p.sarriguren@csic.es}

\date{\today}

\begin{abstract}

  Weak decays in superheavy nuclei with proton numbers $Z=118-120$ and neutron
  numbers $N=175-184$ are studied within a microscopic formalism based on deformed
  self-consistent Skyrme Hartree-Fock mean-field calculations with pairing
  correlations. The half-lives of $\beta^+$ decay and electron capture are compared
  with $\alpha$-decay half-lives obtained from phenomenological formulas.
  The sensitivity of the half-lives to the unknown $Q$-energies is studied by
  comparing the results obtained from different approaches for the masses.
  It is shown that $\alpha$-decay is always dominant in this mass region. The
  competition between $\alpha$ and $\beta^+/EC$ decay modes is studied in seven
  $\alpha$-decay chains starting at different isotopes of $Z$=118, 119, and 120.

\end{abstract}


\end{frontmatter}


\section{Introduction}

The last decades have witnessed a lot of progress in the search and discovery of
increasingly heavy elements and it is nowadays a very fruitful line of research
\cite{hofmann_00,hamilton_13,oganessian_15,hofmann_16,giuliani_19}.
Superheavy nuclei (SHN) with $Z=107-113$ were synthesized from cold-fusion reactions
by using target nuclei $^{208}$Pb and $^{209}$Bi and medium-mass stable isotopes of
Ti, Cr, Fe, Ni, and Zn as projectiles \cite{hofmann_00,hamilton_13,morita_04}.
Production of heavier elements from these reactions were difficult because of the
strong Coulomb repulsion for increasing charge of the projectiles. Then, hot-fusion
reactions involving long-lived actinide nuclei from $^{238}$U to$^{249}$Cf as targets
and the double magic nucleus $^{48}$Ca as projectiles were carried out to produce SHN
with $Z$=112--118 in the neutron-evaporation ($xn$) channels
\cite{hamilton_13,oganessian_04,oganessian_07,oganessian_15_npa}. As a consequence
of these experimental campaigns all the elements with $Z\leq 118$ have been discovered.

However, theoretical macroscopic-microscopic models \cite{nilsson_68,moller_94,kuzmina_12}
that include self-consistent treatments of the shell corrections
\cite{rutz_97,kruppa_00,bender_01,meng_06,agbemava}, predict new regions of particularly
stable nuclear systems with proton shells closures at $Z$ = 114, 120, 124 or 126 and
neutron shell closures at $N$ = 172, 184, depending on the interactions and
parametrizations used. Since no clear indications of closed shell at $Z=114$ or $N=172$
have been observed, there is a strong motivation for the search of more neutron-rich
isotopes, as well as of heavier elements in an attempt to get closer to the predicted
regions of stability.
Concerning superheavy neutron-rich isotopes, alternative ways for their production are
being explored \cite{lopez_19,hessberger_19,hong} through fusion-evaporation reactions
that include not only $xn$ channels, but also the emission of charged particles from the
compound nucleus in the $pxn$ and $\alpha xn$ channels, as well as through multinucleon
transfer reactions or fusion reactions with radioactive ion beams \cite{adamian_epja}.
The production of elements beyond oganesson requires complete fusion reactions with
projectiles with $Z>20$ because of the insufficient amounts of actinide targets with
$Z>98$ available \cite{hofmann_16,adamian_04}.
Different possibilities of projectiles ($^{50}$Ti, $^{54}$Cr, $^{58}$Fe, $^{64}$Ni) and
targets ($^{249}$Cf, $^{248}$Cm, $^{244}$Pu,  $^{238}$U) have been recently studied, both
experimentally and theoretically, in a search for the most suitable combination to
produce elements with $Z>118$. Albers et al. \cite{albers} have measured mass and
angle distributions of fission fragments for several reactions, concluding that
$^{50}$Ti + $^{249}$Cf has the highest fusion probability among the reactions studied
and thus, is the best candidate for the formation of $Z=120$. Similarly, Adamian et
al. \cite{adamian_prc101} found within a microscopic-macroscopic approach that, among
the reactions studied,  $^{50}$Ti + $^{249}$Bk and  $^{50}$Ti + $^{249}$Cf have the
largest cross section for the production of evaporation residues with $Z=119$ and
$Z=120$, respectively.

The stability of the compound nuclei in the superheavy region is generally determined
by spontaneous fission. However, near the predicted islands of stability, fission
barriers increase because of associated effects of shell closures and the half-lives
of spontaneous fission may increase dramatically as shown in Ref. \cite{baran15}.
This enables other radioactive decay modes, such as $\alpha$-decay or weak decays,
that may come into play. In particular, the $\beta^+/EC$-decay in SHN  may open
new pathways towards the predicted region of stability \cite{karpov_12,zagrebaev_12}.
This possibility is also being studied experimentally \cite{hessberger_16,khuyagbaatar}.
Theoretical predictions of weak decays are based on different approaches.
Phenomenological parametrizations \cite{zhang_06} have been developed that can be used
to extrapolate to regions where the $\beta^+/EC$ half-lives are unknown. There are
also calculations that neglect nuclear structure effects, such as those in
Refs. \cite{karpov_12,zagrebaev_12,fiset_72,singh_20}, where only transitions connecting
parent and daughter ground states are considered. The nuclear matrix elements of these
transitions were assumed to be a constant value phenomenologically determined and valid
for all nuclei. However, this value can vary by almost two orders of magnitude
(from $\log(ft)=4.7$ up to $\log(ft)=6.5$), depending on the reference.
In a different approach, half-lives for $\beta^+/EC$-decay were also evaluated within
a proton-neutron quasiparticle random-phase approximation (pnQRPA) based on a
phenomenological folded-Yukawa single-particle Hamiltonian \cite{moller_19}.

Following the work started in Refs. \cite{sarri_19,sarri_20}, we study here
the $\beta^+/EC$-decay half-lives of some selected even-even and odd-$A$ isotopes
with $Z=118-120$ and $N=175-184$ and the competition with $\alpha$-decay.
The production of new elements with $Z=119$ and 120 is one of the main objectives
at worldwide leading laboratories such as SHI-GSI and FLEROV-JINR-DUBNA. Therefore,
the study in this work addresses a highly topical issue.
Furthermore, a comparison between $\alpha$ and $\beta^+/EC$-decay modes is made for
seven $\alpha$-decay chains that follow the production of isotopes with
$Z$=118, 119, and 120.
The method of calculation of the weak decays is based on the pnQRPA approach with
a microscopic nuclear structure calculation consisting on a deformed self-consistent
Hartree-Fock calculation with Skyrme interactions and pairing correlations in the
BCS approximation (HF+BCS).

\section{Theoretical formalism}

The microscopic approach used in this work to calculate $\beta^+/EC$-decay half-lives
is presented here. The method follows closely the theoretical formalism  used in
Ref. \cite{sarri_19,sarri_20} for SHN. Further details of the formalism can be found
elsewhere \cite{sarri1,sarri_form}.

The $\beta^+/EC$-decay half-life, $T_{\beta^+/EC}$, is calculated by summing all the 
allowed Gamow-Teller (GT) transition strengths connecting the parent ground state with
states in the daughter nucleus with excitation energies, $E_{ex}$, lying below 
the $Q_i$ energy ($i=\beta^+,EC$) and weighted with phase-space factors 
$f^i(Z,Q_i-E_{ex})$,

\begin{equation}
T_{1/2,i}^{-1}=\frac{\left( g_{A}/g_{V}\right) _{\rm eff} ^{2}}{D}
\sum_{0 < E_{ex} < Q_i}f^i\left( Z,Q_{i}-E_{ex} \right) B(GT,E_{ex}) \, ,
 \label{t12}
\end{equation}
with $D=6143$ s and $(g_A/g_V)_{\rm eff}=0.77(g_A/g_V)_{\rm free}$, where 0.77 is a 
standard quenching factor and $(g_A/g_V)_{\rm free}=-1.270$.
Therefore, the basic pieces of the calculation are the energies $Q_i$, the
phase-space factors $f^i$ and the GT strength distribution $B(GT,E_{ex})$.

$Q_i$ energies are crucial to evaluate the half-lives because they determine the 
maximum energy of the transition and the values of the phase factors that weight 
the GT strength. They are defined in terms of the nuclear masses $M(A,Z)$ and the
electron mass $m_e$,

\begin{equation}
 Q_{EC}=Q_{\beta^+} +2m_e= M(A,Z)-M(A,Z-1)+m_e \, .
\end{equation}
Usually, one takes experimental masses to evaluate $Q_{EC}$, but in the case of
the SHN studied here the masses have not been determined yet. Therefore, one has
to rely on theoretical predictions for them. There is a large number of mass
formulas available, which have been obtained from different approaches.
We have considered in this work a selection of these mass formulas to evaluate the 
sensitivity of the half-lives to the unknown $Q$-energies. This breaks the
self-consistency of the microscopic calculation, but provides a measure of the
uncertainties involved in the calculated half-lives.

Among the pure phenomenological approaches for the masses, the Weizsacker-Bethe
(WB) nuclear mass formula \cite{wb} is used. Several macroscopic-microscopic
models are also considered. Among them we use the finite-range droplet model
(FRDM) \cite{FRDM}, which is corrected with microscopic effects obtained from a
deformed single-particle model based on folded-Yukawa potentials including pairing
in the Lipkin-Nogami approach and the nuclear mass formula of Ref. \cite{ktuy}
(KTUY) that combines a gross term describing the general trend of the masses,
an even-odd term, and a shell correction term describing the deviations of the
masses from the general trend. The Duflo and Zuker (DZ-10) mass model \cite{DufloZuker},
which is written as an effective Hamiltonian that contains monopole and multipole terms,
is used as well. Another macroscopic-microscopic mass formula inspired by the Skyrme
energy-density functional is also considered. In particular, we use the
Weizsacker-Skyrme formula WS4 that includes a surface diffuseness correction for
unstable nuclei and radial basis function corrections (WS4+RBF) \cite{ws4}.
This mass formula has been shown to be very reliable describing SHN \cite{wang15}.
Finally, we also compare with fully microscopic calculations based on effective
two-body Skyrme nucleon-nucleon interactions by using the masses from the Skyrme
forces SkM* and SLy4 with a zero-range pure volume pairing force \cite{stoitsov}
and Lipkin-Nogami method obtained from the code HFBTHO \cite{mass_sly4}. Tables
for these masses can be found on websites \cite{web_masses}.

\begin{figure}[bth]
\centering
\includegraphics[width=75mm]{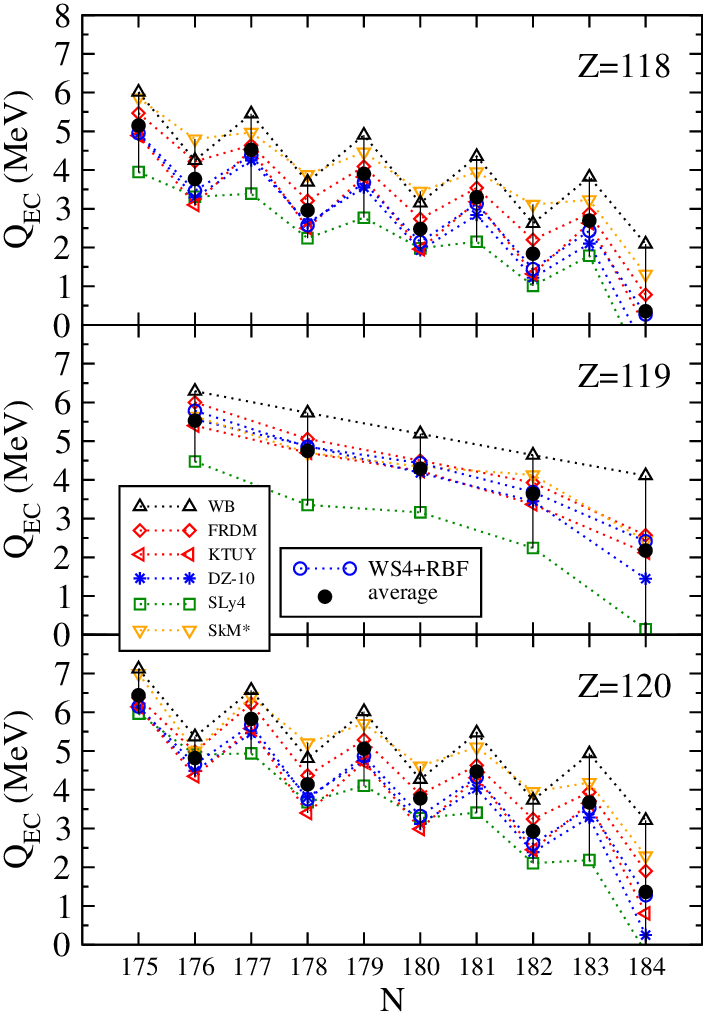}
\caption{$Q_{EC}$ energies (MeV) corresponding to different calculated
masses for $Z=$ 118, 119, and 120 isotopes.}
\label{qbeta}
\end{figure}

\begin{figure}[bth]
\centering
\includegraphics[width=75mm]{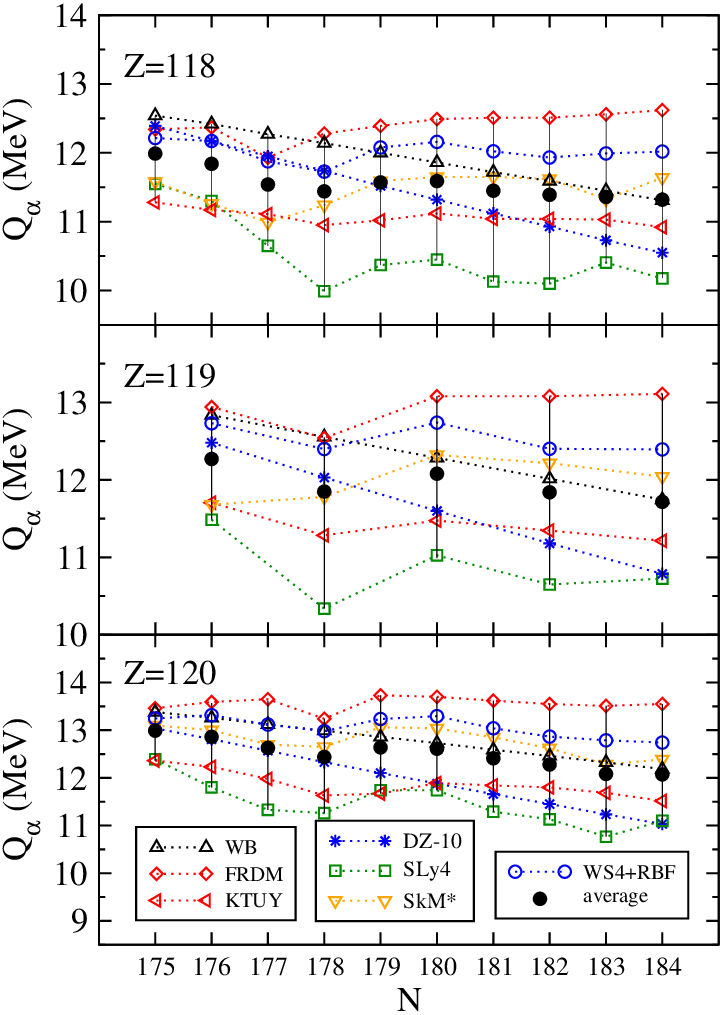}
\caption{Same as in Fig. \ref{qbeta}, but for $Q_{\alpha}$ energies.}
\label{qalfa}
\end{figure}

Figure \ref{qbeta} shows the $Q_{EC}$ energies for the isotopes with $Z=118,119,120$
obtained from the mass formulas mentioned above. It also contains (black solid
circles) the average values for each isotope. The results for each isotope are
typically distributed around 2 MeV. Similarly, Fig. \ref{qalfa} shows the $Q_{\alpha}$
energies, $Q_{\alpha}  = M(A,Z)-M(A-4,Z-2)-M(4,2)$, calculated with the same mass
formulas. They show a similar spread of the results.

The phase-space factors contain two components, positron emission $f^{\beta^+}$ 
and electron capture $f^{EC}$. They are computed numerically for each value of 
the energy using the code LOGFT, as explained in Ref. \cite{gove}.

\begin{equation}
f^{\beta^+} (Z, W_0) = \int^{W_0}_1 p W (W_0 - W)^2 \lambda^+(Z,W) {\rm d}W\, ,
\label{phase}
\end{equation}
with
\begin{equation}
\lambda^+(Z,W) = 2(1+\gamma) (2pR)^{-2(1-\gamma)} e^{-\pi y}
\frac{|\Gamma (\gamma+iy)|^2}{[\Gamma (2\gamma+1)]^2}\, ,
\end{equation}
where $\gamma=\sqrt{1-(\alpha Z)^2}$ ; $y=\alpha ZW/p$ ; $\alpha$ is the fine 
structure constant and $R$ the nuclear radius. $W$ is the total energy of the 
$\beta$ particle, $W_0$ is the total energy available and $p=\sqrt{W^2 -1}$ is
the momentum.

The electron capture phase factors, $f^{EC}$, are given by

\begin{equation}
f^{EC}=\frac{\pi}{2} \sum_{x} q_x^2 g_x^2B_x \, ,
\end{equation}
where $x$ denotes the atomic sub-shell from which the electron is captured that 
includes $K$- and $L$- orbits. $q$ is the neutrino energy, $g$ is the radial 
component of the bound-state electron wave function at the nuclear surface,
and $B$ stands for other exchange and overlap corrections \cite{gove} that
come from the indistinguishability of the electrons and from the decrease of
the nuclear charge by one unit during the decay, respectively.
The bound-state radial wave functions and the correction factors are obtained from a
relativistic self-consistent mean-field calculation. They are solutions of the Dirac
equation with a Hartree self-consistent potential and exchange terms included in the
Slater approximation \cite{martin}. The nuclear potential corresponds to a finite-size
nucleus with a Fermi distribution for the nuclear charge density.

Various improvements have been implemented recently to calculate more accurately the
phase space factors. They have led to a more precise evaluation of the theoretical
half-lives for $\beta$-decay and electron captures \cite{stoica,mougeot}. The final
result found in these works is a small correction of a few percent with respect to
standard calculations in the framework of ref. \cite{gove}. This may be quite important
in some specific cases when comparing with very precise experimental data, but it is
irrelevant in this case, where the purpose is to compare half-lives of different decay
processes that differ by several orders of magnitude, as we shall see later. This
change is also irrelevant when compared to the change induced by other uncertainties
studied in this work, such as the unknown $Q_{EC}$ energies.
Therefore, the use of more accurate electron wave functions will not change the main
conclusion of this work regarding the competition between $\alpha$ and $\beta$
decay modes.

The nuclear structure involved in the $\beta^+/EC$-decay is contained in the energy
distribution of the GT strength $B(GT,E_{ex})$. At variance with other approaches
mentioned earlier to calculate $T_{\beta^+/EC}$ in SHN, we use in this work a
microscopic approach.
We start with a self-consistent calculation of the mean field by means of a deformed
Hartree-Fock procedure with Skyrme interactions and pairing correlations in the BCS
approximation. This calculation provides us single-particle energies, wave functions,
and occupation probabilities.  The Skyrme interaction SLy4 \cite{chabanat} is chosen
for this study because of its proven ability to describe successfully nuclear
properties throughout the entire nuclear chart \cite{stoitsov}. The solution of the
HF equations is found by using the formalism developed in Ref. \cite{vautherin},
under the assumption of time reversal and axial symmetry. The single-particle wave
functions are expanded in terms of the eigenstates of an axially symmetric harmonic
oscillator in cylindrical coordinates using 16 major shells, after verifying that this
size is large enough to get convergence of the HF energies.
Deformation-energy curves (DECs) are constructed by constrained HF calculations that
allow to analyze the nuclear binding energies as a function of the quadrupole
deformation parameter $\beta_2$. The profiles of these curves are found to converge
with the basis used. Furthermore, eventual truncation errors become largely cancelled
out when subtracting energies between the two decay partners to calculate Q-values.

\begin{figure}[bth]
\centering
\includegraphics[width=75mm]{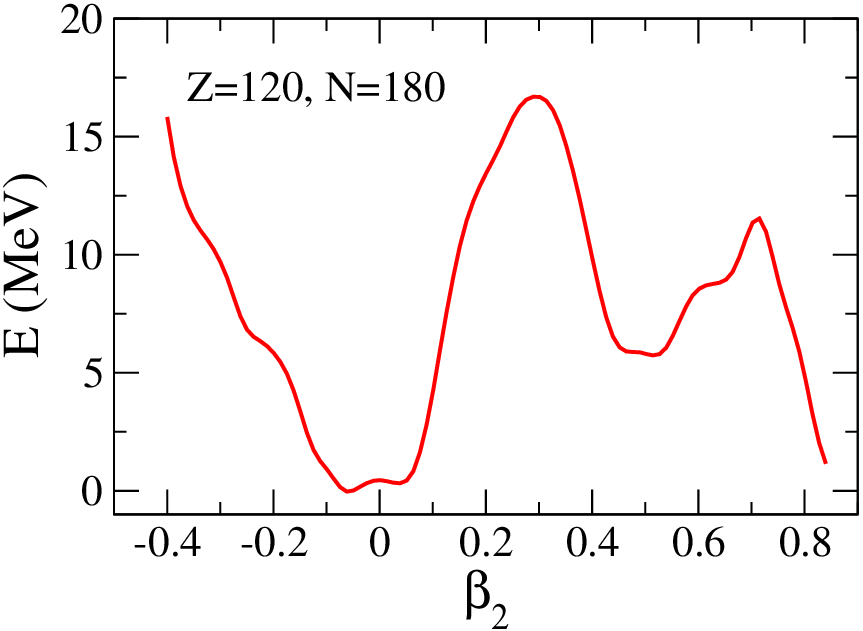}
\caption{Deformation-energy curve for the $^{300}$120 isotope obtained from 
constrained HF+BCS calculations with the Skyrme force SLy4.}
\label{eq0}
\end{figure}


Figure \ref{eq0} shows the DEC of the isotope $^{300}$120 as a representative example
of the nuclei in this mass region. The energy in Fig. \ref{eq0} is relative to the
ground state energy. The results show a ground state corresponding to an almost
spherical shape, as well as an excited prolate minimum at $\beta_2\approx 0.5$. 
The profile of the DEC turns out to be very similar to the DECs obtained for the other
isotopes discussed in this work and agree also quite well with calculations performed
with the finite-range Gogny D1S interaction \cite{gogny}. In this work we calculate
energy distributions of the GT strength and their corresponding half-lives for the
ground state configurations at $\beta_2\approx -0.05$, as well as for the prolate
configuration at $\beta_2\approx 0.5$ that might be populated in the de-excitation of
the compound nucleus. Nuclear deformation has been shown to be a key ingredient to
describe $\beta$-decay properties in many different mass regions \cite{sarri1,sarri_form}
and it is also expected to play a significant role in SHN \cite{sarri_19,sarri_20}.
A deformed pnQRPA with residual spin-isospin interactions is used to obtain the energy
distribution of the GT strength needed to calculate the half-lives. In the case of SHN
the coupling strengths of the residual interactions that scale with the inverse of the
mass number are expected to be very small and their effect is neglected. 

In the case of odd-$A$ nuclei the procedure followed is based on the blocking of a
given state with a given spin and parity, using the equal filling approximation to
calculate its nuclear structure \cite{sarri_form}. This approximation has been shown
to be sufficiently precise for most practical applications \cite{schunck_10}.
The blocked state is chosen among the states in the vicinity of the Fermi level
as the state that minimizes the energy.

\begin{figure}[t]
\centering
\includegraphics[width=75mm]{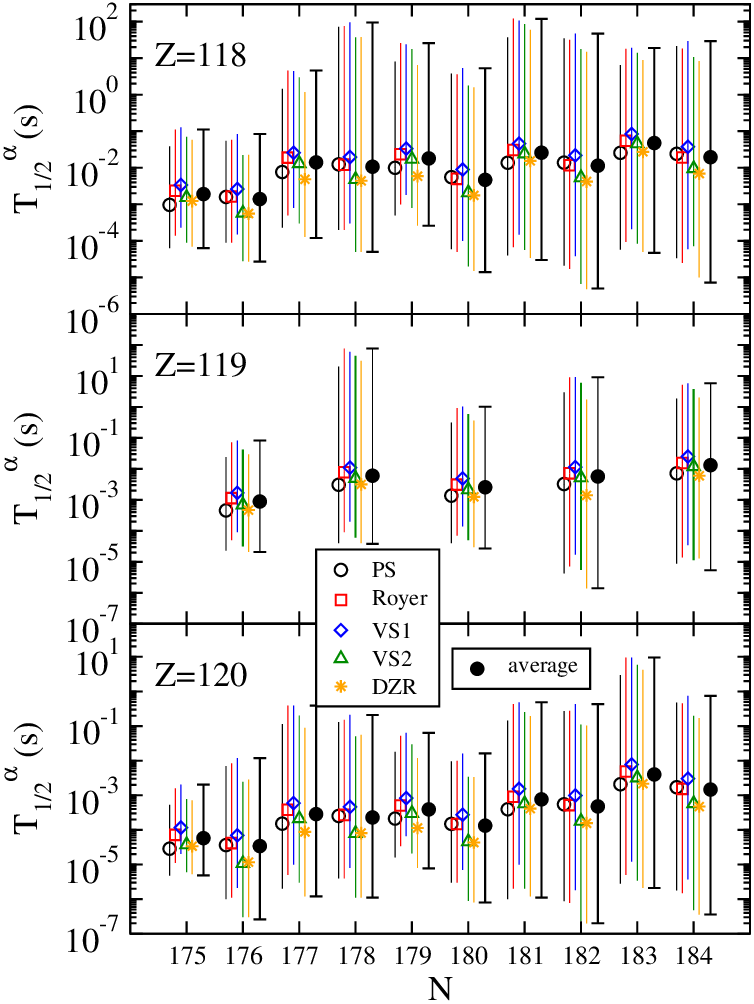}
\caption{Calculated $\alpha$-decay half-lives for $Z=$ 118, 119, and 120 isotopes
  for five different phenomenological formulas of $T_{\alpha}$, namely, PS
  \cite{parkhomenko_05}, Royer \cite{royer_00}, VS1  \cite{viola_66},  
  VS2 \cite{karpov_12,sobiczewski_89}, and  DZR \cite{royer_new} and their average
  value.}
\label{talfa}
\end{figure}

\begin{figure}[t]
\centering
\includegraphics[width=75mm]{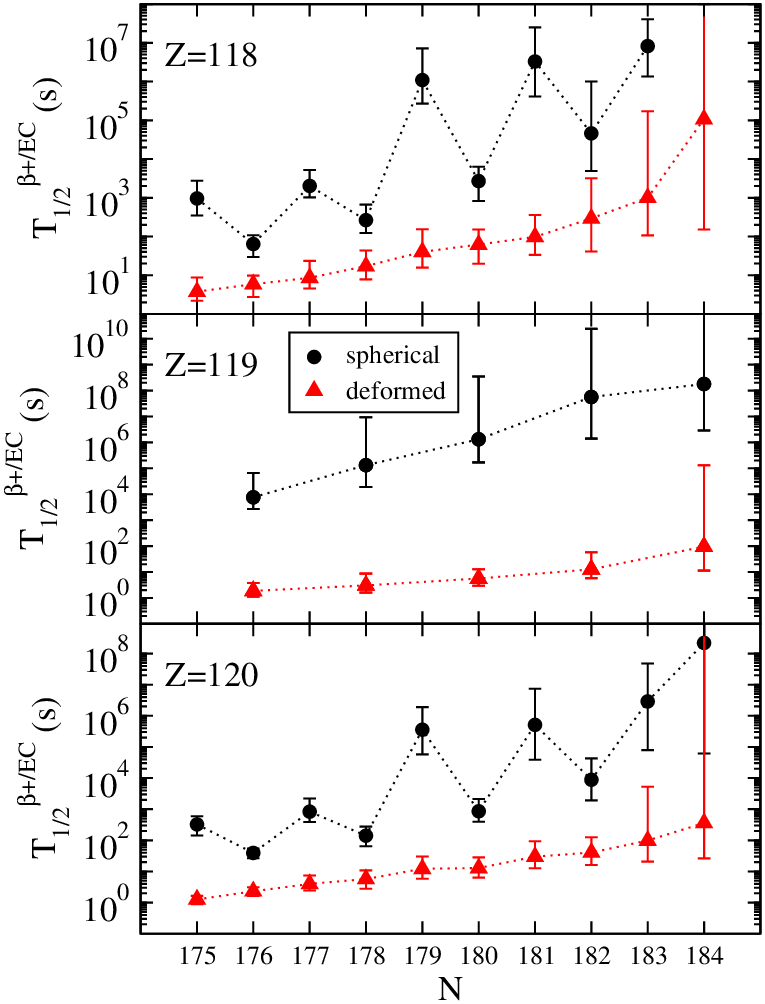}
\caption{Calculated $\beta^+/EC$-decay half-lives for $Z=$ 118, 119, and 120 isotopes,
  using the average, maximum. and minimum $Q_{EC}$ values in Fig. \ref{qbeta}. Results
  for the ground-state spherical and excited prolate deformed shapes are shown.}
\label{tbeta}
\end{figure}


This model of nuclear structure has been successfully used in the past to calculate
weak-decay properties in different mass regions including neutron-deficient medium-mass
\cite{sarri_wp,sarri_rp} and heavy nuclei \cite{sarri_pb1,sarri_pb2,sarri_pb3},
neutron-rich nuclei \cite{sarripere1,sarripere2,sarripere3,kiss,sarri_rare}, and $
fp$-shell nuclei \cite{sarri_fp1,sarri_fp2}.
The effect of various ingredients of the model like deformation and residual 
interactions on the GT strength distributions, which finally determine the 
decay half-lives, was also studied in the above references. In particular, the
sensitivity of the GT distributions to deformation has been used to learn about
the nuclear shapes when comparing with experiment \cite{expnacher}.


\section{Half-lives: Results and discussion}

Comparison between $\alpha$- and $\beta^+/EC$-decay modes is crucial to understand
the possible branching and pathways of the original compound nucleus leading to 
stability.

Since no experimental information is still available on the $\alpha$-decay half-lives
($T_{\alpha}$), one has to rely on phenomenological formulas, which in turn depend on
the unknown $Q_{\alpha}$ values.
Thus, to get an idea of the spread of the results on $T_{\alpha}$ expected from
uncertainties in both $Q_{\alpha}$ energies and phenomenological formulas of
$T_{\alpha}$, we have calculated the half-lives from five parametrizations and 
seven mass formulas, as well as the average, maximum, and minimum  values. 

Following the same approach as in Ref. \cite{sarri_19}, several parametrizations
are used, which were fitted to account for the properties of SHN. Namely, they 
are the formula by Parkhomenko and Sobiczewski \cite{parkhomenko_05} (PS), the
Royer formula \cite{royer_00} (Royer), and the Viola-Seaborg formula \cite{viola_66}
with parameters from \cite{parkhomenko_05} (VS1) and \cite{karpov_12,sobiczewski_89}
(VS2). In addition to these formulas we also consider here a recent formula
\cite{royer_new} (DZR) that takes into account both the blocking effect of the unpaired
nucleon and the contribution of the centrifugal potential. This is an improvement of
the Royer formula, which is simpler and more accurate.

Figure \ref{talfa} contains the results for $T_{\alpha}$. The values shown with
different symbols and colors correspond to calculations from the five different
parametrizations using the average values of $Q_{\alpha}$ (see Fig. \ref{qalfa}).
The error bars for each calculation correspond to the use of the maximum and
minimum $Q_{\alpha}$ values predicted by the different mass evaluations.
Solid black circles correspond to the average values of the five formulas. Their
vertical lines join the maximum and minimum values obtained from the different
$T_{\alpha}(Q_{\alpha})$ alternatives. While the predictions of different parametrizations
are within one order of magnitude, the uncertainties originated from the  $Q_{\alpha}$
energies may vary up to seven orders of magnitude.

Figure \ref{tbeta} contains the results for $T_{\beta^+/EC}$ calculated with the
formalism described in the previous section. The results plotted with a symbol
(circle or triangle) correspond to the use of the average values of $Q_{EC}$ in
Fig. \ref{qbeta}. The maximum and minimum values of $T_{\beta^+/EC}$ for each
isotope correspond to the minimum and maximum values of $Q_{EC}$, respectively,
according to the different mass formulas. They are plotted as error bars that join
these extreme values, giving a measure of the uncertainties associated with the
unknown $Q_{EC}$ energies.

The results with black circles correspond to the slightly oblate, almost spherical,
ground states ($\beta_2\approx 0$), whereas the red triangles stand for the results
from the deformed prolate shapes ($\beta_2\approx 0.5$). Clearly, the half-lives from
deformed shapes are much lower than those from spherical shapes for a given isotope.
The reason of this behavior can be traced back to the different scenarios in the
deformed or spherical cases having very different density of levels.
In the $\beta^+/EC$-decay one proton is transformed into one neutron. The low-lying
excitations below $Q_{EC}$ in the daughter nucleus that finally determine the half-lives
come from transitions connecting protons around the Fermi level for $Z=118-120$ with
neutrons around the Fermi level in the mass region $N=175-184$.
In a pure spherical approach, the GT operator should match protons from the spherical
shell $2f_{5/2}$ of negative parity with neutrons from the
$2g_{7/2},\, 3d_{3/2},\, 3d_{5/2},\, 4s_{1/2}$ shells, which are positive parity states.
Thus, these transitions will be very suppressed in nuclei with small deformations
because of parity arguments.
On the other hand, in the deformed case, many different orbitals from different
spherical shells with positive and negative parity cross each other leading to a
much more mixed scenario where states with both parities are found in the vicinity
of the Fermi levels of protons and neutrons.
The final result is an enhancement of the  GT strength at low
energy that leads to a shorter half-life in the case of deformed nuclei.

Another interesting observation is the existence of an odd-even staggering effect
in the spherical case, which does not appear in the deformed case. This peculiar
behavior is related to the characteristics of the excitations in the odd nuclei.
The low-lying transitions in the odd system correspond basically to
one-quasiparticle (1qp) excitations where the odd nucleon is involved in the process.
At higher excitation energies, typically beyond the energy needed to break a pair
of nucleons, the transitions are mainly three-quasiparticle (3qp) excitations similar
to those in the even-even system but with the odd nucleon acting as a spectator.
For nuclei in this mass region, which have rather small $Q_{EC}$ energies, the 3qp
excitations are shifted in energy beyond $Q_{EC}$, while the low-lying 1qp excitations
connecting protons from the $2f_{5/2}$ shell with neutrons in the 
$2g_{7/2},\, 3d_{3/2},\, 3d_{5/2},\, 4s_{1/2}$ shells are very suppressed because of
parity. The final result for spherical nuclei is that in the odd nuclei very little
strength remains within the $Q_{EC}$ window giving rise to quite large half-lives as
compared with the even-even nuclei.
In the deformed case this effect is not manifest because of the higher level density
around the Fermi levels that involve states with both positive and negative
parities, as well as many angular momentum components.

Comparing the half-lives in Figs. \ref{talfa} and \ref{tbeta}, one can see that the
$\beta^+/EC$-decay half-lives are systematically several orders of magnitude larger
than the corresponding average $\alpha$-decay half-lives for a given isotope.
The range of this difference is between three and five orders of magnitude in the
deformed case and even larger in the spherical one. Only when one considers
the maximum values of $T_\alpha$ allowed by the uncertain $Q_\alpha$ energies are then
comparable to $T_{\beta^+/EC}$ in the deformed case.
As a consequence, $\alpha$-decay in this mass region will be always dominant and
much faster than $\beta^+/EC$-decay.

Finally, Tables \ref{table1}--\ref{table3} show the $Q_\alpha$ and $Q_{EC}$ energies,
as well as $T_\alpha$ and $T_{\beta^+/EC}$ half-lives for nuclei involved in various
$\alpha$-decay chains starting at $^{295}$Og and $^{296}$Og (Table \ref{table1}),
$^{295}$119 (Table \ref{table2}), and $^{295}$120, $^{296}$120, $^{297}$120, and
$^{298}$120 (Table \ref{table3}). The last column stands for the ratios
${\cal R}=T_{\beta^+/EC}/T_{\alpha}$.

Since the SHN produced after neutron evaporation of the corresponding compound
nuclei are identified by their $\alpha$-decay chains, the competition between
$\alpha$ and $\beta^+/EC$ decay modes in the members of a given chain of
$\alpha$-decays is important to analyze possible branching points in future
experiments.


\begin{table}[h!]
  \caption{$Q_\alpha$ and $Q_{EC}$ energies [MeV], as well as $T_\alpha$  and
    $T_{\beta^+/EC}$ half-lives [s] for nuclei in the $\alpha$-decay chains starting
    at  $^{295}$Og and $^{296}$Og. Available experimental values appear with the
    errors within parentheses. Otherwise, they are calculated values (see text).
    The last column contains the ratio ${\cal R}=T_{\beta^+/EC}/T_{\alpha}$}.
  \centering
           \scalebox{0.8}{
      {\begin{tabular}{cccccc} \hline \\
          & $Q_\alpha$  & $Q_{EC}$& $T_\alpha$ & $T_{\beta^+/EC}$ & ${\cal R}$ \\
          \hline  \\
 $^{295}$Og  &  11.7(9)   & 4.38     & 7.1$\times 10^{-3}$     & 2.3$\times 10^{3}$   & 3.2$\times 10^{5}$ \\
 $^{291}$Lv  &  10.89(5)  & 3.4(10)  & 2.8$\times 10^{-2}$(15) & 8.1$\times 10^{2}$   & 2.9$\times 10^{4}$ \\
 $^{287}$Fl  &  10.16(5)  & 2.83(95) & 5.2$\times 10^{-1}$(13) & 1.1$\times 10^{3}$   & 2.1$\times 10^{3}$ \\
 $^{283}$Cn  &  9.94(11)  & 2.21(93) & 4.1$\times 10^{0}$(10)  & 2.2$\times 10^{3}$   & 5.4$\times 10^{2}$ \\
 $^{279}$Ds  &  10.08(11) & 1.63(90) & 2.6$\times 10^{-1}$     & 7.6$\times 10^{3}$   & 2.9$\times 10^{4}$ \\
 $^{275}$Hs  &  9.44(5)   & 0.93(84) & 2.9$\times 10^{-1}$(15) & 4.4$\times 10^{4}$   & 1.5$\times 10^{5}$ \\
 $^{271}$Sg  &  8.89(11)  & 0.16     & 2.9$\times 10^{1}$      & -                 & - \\
 $^{267}$Rf  &  7.89(30)  & -0.48    & 1.2$\times 10^{4}$      & -                 & - \\
          \hline \\
          
  $^{296}$Og  &  11.73      & 2.58 & 2.3$\times 10^{-3}$     & 4.2$\times 10^{2}$ & 1.8$\times 10^{5}$ \\
  $^{292}$Lv  &  10.774(15) & 2.33 & 2.4$\times 10^{-2}$(12) & 2.7$\times 10^{3}$ & 1.1$\times 10^{5}$ \\
  $^{288}$Fl  &  10.072(13) & 1.14 & 7.5$\times 10^{-1}$(14) & 8.0$\times 10^{6}$ & 1.1$\times 10^{7}$ \\
  $^{284}$Cn  &  9.60(20)   & 0.84 & 8.1$\times 10^{0}$      & 6.2$\times 10^{6}$ & 7.7$\times 10^{5}$ \\
\hline 
\label{table1}
\end{tabular}}}
\end{table}


\begin{table}[h!]
  \caption{Same as in Table \ref{table1}, but for the $\alpha$-chain starting at
    $^{295}$119.}
  \centering
         \scalebox{0.8}{
     {\begin{tabular}{cccccc}  \hline \\
         & $Q_\alpha$  & $Q_{EC}$& $T_\alpha$ & $T_{\beta^+/EC}$ & ${\cal R}$ \\
         \hline \\
 $^{295}$119 & 12.73     & 5.79     & 9.1$\times 10^{-5}$     & 4.8$\times 10^{3}$ & 5.3$\times 10^{7}$ \\ 
 $^{291}$Ts  & 11.5(4)   & 4.41(85) & 1.2$\times 10^{-2}$     & 3.8$\times 10^{4}$ & 3.2$\times 10^{6}$ \\ 
 $^{287}$Mc  & 10.76(5)  & 3.82(75) & 9.5$\times 10^{-2}$(60) & 3.1$\times 10^{3}$ & 3.3$\times 10^{4}$ \\
 $^{283}$Nh  & 10.51(11) & 3.22(75) & 1.6$\times 10^{-1}$(10) & 5.3$\times 10^{1}$ & 3.3$\times 10^{2}$ \\
 $^{279}$Rg  & 10.52(5)  & 2.65(73) & 1.8$\times 10^{-1}$(11) & 1.1$\times 10^{2}$ & 5.9$\times 10^{2}$ \\
 $^{275}$Mt  & 10.48(5)  & 2.21(72) & 1.17$\times 10^{-1}$(74)& 5.1$\times 10^{3}$ & 4.4$\times 10^{4}$ \\
 $^{271}$Bh  & 9.42(5)   & 1.16(72) & 1.7$\times 10^{0}$      & 1.4$\times 10^{5}$ & 8.2$\times 10^{4}$ \\
 $^{267}$Db  & 7.92(30)  & 0.63(71) & 2.8$\times 10^{4}$      & 6.2$\times 10^{3}$ & 2.2$\times 10^{-1}$ \\
 $^{263}$Lr  & 7.68(20)  & 0.60(57) & 3.1$\times 10^{4}$      & 5.0$\times 10^{4}$ & 1.6$\times 10^{0}$ \\ 
\hline \\
\label{table2}
\end{tabular}}}
\end{table}


\begin{table}[h!]
  \caption{Same as in Table \ref{table1}, but for the $\alpha$-chains starting at
    $^{295}$120, $^{296}$120, $^{297}$120, and $^{298}$120.}
  \centering
       \scalebox{0.8}{
  {\begin{tabular}{cccccc} \hline  \\
      & $Q_\alpha$  & $Q_{EC}$  & $T_\alpha$ & $T_{\beta^+/EC}$ & ${\cal R}$ \\
      \hline \\
 $^{295}$120 & 13.25     & 6.15     & 1.7$\times 10^{-5}$ & 4.6$\times 10^{2}$ & 2.7$\times 10^{7}$ \\
 $^{291}$Og  & 12.39     & 5.63     & 2.5$\times 10^{-4}$ & 9.8$\times 10^{1}$ & 3.9$\times 10^{5}$ \\
 $^{287}$Lv  & 11.25     & 4.93     & 2.5$\times 10^{-2}$ & 7.9$\times 10^{1}$ & 3.1$\times 10^{3}$ \\
 $^{283}$Fl  & 10.84     & 4.15     & 5.8$\times 10^{-2}$ & 2.4$\times 10^{1}$ & 4.2$\times 10^{2}$ \\
 $^{279}$Cn  & 11.04(20) & 3.26(62) & 4.5$\times 10^{-3}$ & 1.7$\times 10^{2}$ & 3.8$\times 10^{4}$ \\
 $^{275}$Ds  & 11.40(30) & 2.74(59) & 1.0$\times 10^{-2}$ & 8.3$\times 10^{2}$ & 8.3$\times 10^{4}$ \\
 $^{271}$Hs  & 9.51(11)  & 1.82(50) & 2.1$\times 10^{0}$  & 5.3$\times 10^{3}$ & 2.5$\times 10^{3}$ \\ 
 $^{267}$Sg  & 8.63(21)  & 1.73(49) & 2.1$\times 10^{2}$  & 2.7$\times 10^{2}$ & 1.3$\times 10^{0}$ \\
 $^{263}$Rf  & 8.25(15)  & 1.03(32) & 6.7$\times 10^{2}$  & 2.3$\times 10^{5}$ & 3.4$\times 10^{2}$ \\
      \hline
      \\
  $^{296}$120 &  13.32    & 4.66     & 4.6$\times 10^{-6}$     & 4.4$\times 10^{1}$  & 9.5$\times 10^{6}$ \\
  $^{292}$Og  &  12.21    & 3.79     & 5.0$\times 10^{-4}$     & 1.9$\times 10^{2}$  & 3.9$\times 10^{5}$ \\
  $^{288}$Lv  &  11.26    & 3.30     & 7.4$\times 10^{-3}$     & 1.2$\times 10^{2}$  & 1.6$\times 10^{4}$ \\
  $^{284}$Fl  &  10.83(3) & 2.33(85) & 3.3$\times 10^{-3}$(14) & 2.0$\times 10^{2}$  & 5.9$\times 10^{4}$ \\
      \hline
      \\
 $^{297}$120 & 13.12     & 5.67      & 2.8$\times 10^{-5}$     & 9.6$\times 10^{2}$ & 3.4$\times 10^{7}$ \\
 $^{293}$Og  & 11.9(9)   & 4.49(107) & 2.9$\times 10^{-3}$     & 1.6$\times 10^{3}$ & 5.5$\times 10^{5}$ \\    
 $^{289}$Lv  & 11.1(3)   & 3.86(95)  & 5.6$\times 10^{-2}$     & 1.3$\times 10^{3}$ & 2.3$\times 10^{4}$ \\
 $^{285}$Fl  & 10.56(5)  & 3.27(90)  & 2.1$\times 10^{-1}$(10) & 6.7$\times 10^{2}$ & 3.2$\times 10^{3}$ \\
 $^{281}$Cn  & 10.45(5)  & 2.72(89)  & 1.8$\times 10^{-1}$(8)  & 5.7$\times 10^{2}$ & 3.2$\times 10^{3}$ \\
 $^{277}$Ds  & 10.83(11) & 2.17(80)  & 6$\times 10^{-3}$(3)    & 5.0$\times 10^{2}$ & 8.3$\times 10^{4}$ \\
 $^{273}$Hs  & 9,70(5)   & 1.26(78)  & 1.06$\times 10^{0}$(50) & 1.4$\times 10^{4}$ & 1.3$\times 10^{4}$ \\
 $^{269}$Sg  & 8.65(5)   & 0.61(72)  & 3.0$\times 10^{2}$(18)  & 1.6$\times 10^{5}$ & 5.3$\times 10^{2}$ \\
 $^{265}$Rf  & 7.81(30)  & 0.46(66)  & 2.5$\times 10^{4}$      & 2.8$\times 10^{4}$ & 1.1$\times 10^{0}$ \\
      \hline
      \\
  $^{298}$120 &  12.98    & 3.76     & 2.0$\times 10^{-5}$    & 2.0$\times 10^{2}$ & 9.8$\times 10^{6}$  \\
  $^{294}$Og  &  11.8(7)  & 2.94(94) & 1.2$\times 10^{-3}$(5) & 1.3$\times 10^{2}$ & 1.1$\times 10^{5}$  \\
  $^{290}$Lv  &  11.00(7) & 2.30(93) & 8$\times 10^{-3}$(3)   & 2.9$\times 10^{3}$ & 3.6$\times 10^{5}$  \\
  $^{286}$Fl  &  10.37(3) & 1.76(93) & 2.8$\times 10^{-1}$    & 4.8$\times 10^{3}$ & 1.7$\times 10^{4}$ \\
\hline 
\label{table3}
      \end{tabular}}}
\end{table}


$Q$-energies in the tables are taken from experiment \cite{audi} when available,
together with their errors within parentheses. Otherwise, they are calculated
from the mass formula WS4-RBF \cite{ws4} and quoted without errors. Similarly,
the half-lives are either experimental with errors or calculated with the
$Q$-values given in the tables. In the latter case, $T_\alpha$ corresponds to
the average value of the phenomenological parametrizations used earlier, whereas
$T_{\beta^+/EC}$ is calculated with the HF+BCS+pnQRPA formalism for the ground states
of the nuclei.
As it can be seen from the tables, the $\alpha$-decay mode of a given isotope is
generally orders of magnitude faster than the corresponding $\beta^+/EC$ decay with
ratios ${\cal R}$ of several orders of magnitude. Nevertheless, ${\cal R}$ decreases
as we progress in odd-$A$ chains and can reach values close to one at the end of some
of these chains. This is the case of nuclei such as $^{267}$Sg, $^{265}$Rf, $^{267}$Db,
and $^{263}$Lr. However, spontaneous fission becomes the dominant decay mode for
them. It is also worth noting that in cases where $Q_{EC}$ is very small, the large
$T_{\beta^+/EC}$ obtained are very sensitive to fine details of the calculations
because they are determined from the few low-lying transitions located just below
$Q_{EC}$.

As we go forward in a given chain, there is a general trend of decreasing values
of the $Q_{\alpha}$ and $Q_{EC}$ energies, which is translated into increasing
half-lives and decreasing ratios. This general trend is somewhat altered
in the vicinity of $Z=112$ and $N=168$ nuclei, where sub-shell effects are expected
at slightly prolate deformations, which are typical of nuclei in this mass
region \cite{sarri_20}.

\section{Conclusions}

$\beta^+/EC$-decay half-lives of some selected even-even and odd-$A$ isotopes in
the region of SHN with proton numbers $Z=118-120$ and neutron numbers $N=175-184$
have been studied. $Z=$119 and 120 are the next elements to be discovered and their
production is object of very active experimental campaigns.

The nuclear structure of the decay partners is described microscopically within a
pnQRPA based on a self-consistent deformed Skyrme HF+BCS approach.
Uncertainties in the $Q$-energies originated from the unknown masses are translated
into uncertainties in the half-lives calculated with them. We have used different
mass formulas to evaluate the expected spread on the half-lives. The results for
$T_{\beta^+/EC}$ are compared with those of $T_{\alpha}$ obtained from several
phenomenological formulas using $Q_{\alpha}$ energies obtained from the same mass
formulas. The $T_{\alpha}$ half-lives are systematically lower than the corresponding
$T_{\beta^+/EC}$ ones for a given isotope. The difference is always larger than three
orders of magnitude in the most favored case of deformed nuclei and becomes much
larger for the spherical ground states.
Therefore, the  $\beta^+/EC$-decay mode will hardly compete
with $\alpha$-decay in the SHN studied with $Z$=118, 119, and 120.

The competition between $\alpha$ and $\beta^+/EC$ decay modes is also studied in seven
$\alpha$-decay chains starting at different isotopes of $Z$=118, 119, and 120.
The ratio of half-lives for both modes indicates that $\alpha$-decay is generally
several orders of magnitude (up to seven orders in the chain heads) faster than
$\beta^+/EC$-decay. However, the half-lives become comparable at the end of some
of the chains studied. Hence, the cases in which different decay branches are more
likely to appear have been identified. This could be useful as a theoretical guide
for future experimental studies of these decays in new elements not yet discovered.

\section*{Acknowledgments}
I would like to thank G. G. Adamian for useful discussions and valuable advice. 
This work was supported by Ministerio de Ciencia e Innovaci\'on 
MCI/AEI/FEDER,UE (Spain) under Contract No. PGC2018-093636-B-I00.  



\section*{References}

\begin{thebibliography}{99}

\bibitem{hofmann_00} S. Hofmann, G. M\"unzenberg, Rev. Mod. Phys. 72 (2000) 733.

\bibitem{hamilton_13} J.H. Hamilton, D. Hofmann, Y.T. Oganessian, 
Annu. Rev. Nucl. Part. Sci. 63 (2013) 383.

\bibitem{oganessian_15} Yu. Ts. Oganessian, V.K. Utyonkov,
Rep. Prog. Phys. 78 (2015) 036301.

\bibitem{hofmann_16} S. Hofmann, et al., Eur. Phys. J. A 52 (2016) 180.

\bibitem{giuliani_19} S.A. Giuliani, et al., Rev. Mod. Phys. 91 (2019) 011001.

\bibitem{morita_04} K. Morita, et al., J. Phys. Soc. Japan 73(10) (2004) 2593.

\bibitem{oganessian_04} Yu. Ts. Oganessian, et al., Phys. Rev. C 69 (2004) 021601(R).

\bibitem{oganessian_07} Yuri Oganessian, J. Phys. G: Nucl. Part. Phys. 34 (2007) R165.
  
\bibitem{oganessian_15_npa} Yu. Ts. Oganessian, V.K. Utyonkov, 
  Nucl. Phys. A 944 (2015) 62.
  
\bibitem{nilsson_68} S.G. Nilsson,  et al., Nucl. Phys. A 115 (1968) 545.

\bibitem{moller_94} P. M\"oller, J.R. Nix, J. Phys. G: Nucl. Part. Phys.  20 (1994) 1681.

\bibitem{kuzmina_12}  A.N. Kuzmina, G.G. Adamian, N.V. Antonenko, W. Scheid, 
Phys. Rev. C 85 (2012) 014319.

\bibitem{rutz_97} K. Rutz, et al., Phys. Rev. C 56 (1997) 238.

\bibitem{kruppa_00} A.T. Kruppa, M. Bender, W. Nazarewicz, P.-G. Reinhard, T. Vertse, 
S. \'Cwiok, Phys. Rev. C 61 (2000) 034313.

\bibitem{bender_01} M. Bender, W. Nazarewicz, P.-G. Reinhard, 
Phys. Lett. B 515 (2001) 42.

\bibitem{meng_06} J. Meng, H. Toki, S.G. Zhou, S.Q. Zhang, W.H. Long, L.S. Geng,
  Prog. Part. Nucl. Phys. 57 (2006) 470.
  
\bibitem{agbemava} S.E. Agbemava, A.V. Afnasjev, T. Nakatsukasa, P. Ring,
  Phys. Rev. C 92 (2015) 054310.
  
\bibitem{lopez_19} A. Lopez-Martens, et al., Phys. Lett. B 795 (2019) 271.

\bibitem{hessberger_19} F.P. He$\beta$berger , Eur. Phys. J. A 55 (2019) 208.

\bibitem{hong} Juhee Hong, G.G. Adamian, N.V. Antonenko,
 Phys. Rev. C 94 (2016) 044606;   Phys. Lett. B 764 (2017) 42.
 
\bibitem{adamian_epja} G.G. Adamian, N.V. Antonenko, A. Diaz-Torres, S. Heinz.
  Eur. Phys. J. A 56 (2020) 47.
  
\bibitem{adamian_04} G.G. Adamian, N.V. Antonenko, W. Scheid,
Phys. Rev. C 69 (2004) 044601.
  
\bibitem{albers} H.M. Albers, et al., Phys. Lett. B 808 (2020) 135626.

\bibitem{adamian_prc101} G.G. Adamian, N.V. Antonenko, H. Lenske, L.A. Malov,
  Phys. Rev. C 101 (2020) 034301.

\bibitem{baran15} A. Baran, M. Kowal, P.-G. Reinhard, L.M. Robledo, A. Staszczak, and
  M. Warda, Nucl. Phys. A 944 (2015) 442.

\bibitem{karpov_12} A.V. Karpov, V.I. Zagrebaev, Y. Martinez Palazuela, L. Felipe Ruiz,
Walter Greiner, Int. J. Mod. Phys. E 21 (2012) 1250013.

\bibitem{zagrebaev_12} V.I. Zagrebaev, A.V. Karpov, Walter Greiner, 
Phys. Rev. C 85 (2012) 014608.

\bibitem{hessberger_16} F.P. He$\beta$berger, et al., Eur. Phys. J. A 52 (2016) 328.

\bibitem{khuyagbaatar} J. Khuyagbaatar, et al., Phys. Rev. Lett. 125 (2020) 142504.

\bibitem{zhang_06} X. Zhang, Z. Ren, Phys. Rev. C 73 (2006) 014305.
  
\bibitem{fiset_72} E.O. Fiset, J.R. Nix, Nucl. Phys. A 193 (1972) 647.
  
\bibitem{singh_20} U.K. Singh, P.K. Sharma, M. Kaushik, S.K. Jain, Dashty T. Akraway,  G. Saxena, Nucl. Pĥys. A 1004 (2020) 122035.

\bibitem{moller_19}  P. M\"oller, M.R. Mumpower, T. Kawano, W.D. Myers, 
At. Data Nucl. Data Tables 125 (2019) 1.

\bibitem{sarri_19} P. Sarriguren, Phys. Rev. C 100 (2019) 014309.
  
\bibitem{sarri_20} P. Sarriguren, J. Phys. G: Nucl. Part. Phys. 47 (2020) 125107.

\bibitem{sarri1}  P. Sarriguren, E. Moya de Guerra, A. Escuderos, A. C. Carrizo, 
Nucl. Phys. A 635 (1998) 55.

\bibitem{sarri_form} P. Sarriguren, E. Moya de Guerra, A. Escuderos, 
Nucl. Phys. A 658 (1999) 13;
Nucl. Phys. A 691 (2001)  631;
Phys. Rev. C 64 (2001) 064306.

\bibitem{wb} C.F. von Weizsacker, Z. Phys. 96 (1935) 431 (1935);
  H.A. Bethe, R.F.Bacher, Rev. Mod. Phys. 8 (1936) 829.

\bibitem{FRDM} P. M\"oller, A.J. Sierk, T. Ichikawa, H. Sagawa, 
  At. Data Nucl. Data Tables 109-110 (2016) 1.

\bibitem{ktuy} H. Koura, T. Tachibana, M. Unos, N. Yamada,
  Prog. Theor. Phys. 113 (2005) 305.
  
\bibitem{DufloZuker} J. Duflo, A.P. Zuker, Phys. Rev. C 52 (1995) R23;
  J. Mendoza-Temis, J.G. Hirsch, A.P. Zuker, Nucl. Phys. A 843 (2010) 14.
  
\bibitem{ws4} N. Wang, M. Liu, X.Z. Wu, J. Meng, Phys. Lett. B 734 (2014) 215.
  
\bibitem{wang15} Y.Z. Wang, S.J. Wang, Z.Y. Hou, J.Z. Gu, Phys. Rev. C 92 (2015) 064301.

\bibitem{stoitsov} M.V. Stoitsov, J. Dobaczewski, W. Nazarewicz, S. Pittel, D.J. Dean,
  Phys. Rev. C 68 (2003) 054312.

\bibitem{mass_sly4} M. V. Stoitsov, J. Dobaczewski, W. Nazarewicz, P. Ring, 
  Comp. Phys. Comm. 167 (2005) 43.

\bibitem{web_masses} https://www.fuw.edu.pl/$\sim$dobaczew/thodri/thodri.html; \\
www.nuclearmasses.org

\bibitem{gove} N.B. Gove, M.J. Martin, Nucl. Data Tables 10 (1971) 205.

\bibitem{martin} M.J. Martin and P.H. Blichert-Toft, Nuclear Data Tables A 8 (1970) 1.

\bibitem{stoica} S. Stoica, M. Mirea, O. Niuescu, J.U. Nabi, and M. Ishfaq,
  Adv. High En. Phys. 2016 (2016) 8729893; \\
  M. Ishfaq, J.U. Nabi, O. Niuescu, M. Mirea, and S. Stoica, Adv. High En. Phys.
  2019 (2019) 5783618.

\bibitem{mougeot} X. Mougeot, Appl. Radiat. Isot. 134 (2018) 225; 154 (2019) 108884.

\bibitem{chabanat} E. Chabanat, P. Bonche, P. Haensel, J. Meyer, R. Schaeffer, 
Nucl. Phys. A 635 (1998) 231.

\bibitem{vautherin} D. Vautherin, D.M. Brink, Phys. Rev. C 5 (172) 626; 
D. Vautherin, Phys. Rev. C 7 (1973) 296.

\bibitem{gogny} S. Hilaire, M. Girod, Eur. Phys. J. A 33 (2007) 237; \\
www-phynu.cea.fr/science\_en\_ligne/carte\_potentiels\_mi\-croscopiques/carte\_potentiel\_nucleaire\_eng.htm

\bibitem{schunck_10} N. Schunck, et al., Phys. Rev. C 81 (2010) 024316.

\bibitem{sarri_wp} P. Sarriguren, R. Alvarez-Rodr\'{\i}guez, E. Moya de Guerra, 
Eur. Phys. J. A 24 (2005) 193.

\bibitem{sarri_rp} P. Sarriguren, Phys. Rev. C 79 (2009) 044315; 
Phys. Lett. B 680 (2009) 438; 
Phys. Rev. C 83 (2011) 025801.

\bibitem{sarri_pb1} P. Sarriguren, O. Moreno, R. Alvarez-Rodr\'{\i}guez,
  E. Moya de Guerra, Phys. Rev. C 72 (2005) 054317.

\bibitem{sarri_pb2} O. Moreno, P. Sarriguren, R. Alvarez-Rodr\'{\i}guez,
  E. Moya de Guerra, Phys. Rev. C 73 (2006) 054302.

\bibitem{sarri_pb3} J.M. Boillos, P. Sarriguren, Phys. Rev. C 91 (2015) 034311.

\bibitem{sarripere1} P. Sarriguren, J. Pereira, Phys. Rev. C  81 (2010) 064314.

\bibitem{sarripere2} P. Sarriguren, A. Algora, J. Pereira, 
Phys. Rev. C 89 (2014) 034311.

\bibitem{sarripere3} P. Sarriguren, Phys. Rev. C 91 (2015) 044304.

\bibitem{kiss} P. Sarriguren, A. Algora, G. Kiss,  
Phys. Rev. C 98 (2018) 024311.

\bibitem{sarri_rare} P. Sarriguren, Phys. Rev. C 95 (2017) 014304.

\bibitem {sarri_fp1} P. Sarriguren, E. Moya de Guerra, R. Alvarez-Rodr\'{\i}guez, 
Nucl. Phys. A 716 (2003) 230.

\bibitem{sarri_fp2} P. Sarriguren, Phys. Rev. C 87 (2013) 045801; 
Phys. Rev. C 93 (2016) 054309.

\bibitem{expnacher} E. N\'acher, et al., Phys. Rev. Lett. 92 (2004) 232501.

\bibitem{parkhomenko_05} A. Parkhomenko, A. Sobiczewski, 
  Acta Physica Polonica B 36 (2005) 3095.

\bibitem{royer_00} G. Royer, J. Phys. G: Nucl. Part. Phys.  26 (2000) 1149.

\bibitem{viola_66} V.E. Viola, Jr., G.T. Seaborg, 
J. Inorg. Nucl. Chem. 28 (1966) 741.

\bibitem{sobiczewski_89} A. Sobiczewski, Z. Patyk, S. \'Cwiok, 
Phys. Lett. B 224 (1989) 1.

\bibitem{royer_new} Jun-Gang Deng, Hong-Fei Zhang, G. Royer,
  Phys. Rev. C 101 (2020) 034307.

\bibitem{audi}  G. Audi, F.G. Kondev, M. Wang, W.J. Huang, S. Naimi,
  Chinese Physics C 41 (2017) 030001; M. Wang. G. Audi, F.G. Kondev,
  W.J. Huang, S. Naimi, X. Xu, Chinese Physics C 41 (2017) 030003.

\end{thebibliography}
\end{document}